\documentclass{aa}
\def\a{\mbox{AE\,Aqr}\ }

\def\e{et~al.\ }

\def\r{R_{\rm m}}
\newcommand{\be}{\begin{equation}}
\newcommand{\ee}{\end{equation}}
\newcommand{\bdm}{\begin{displaymath}}
\newcommand{\edm}{\end{displaymath}}

\usepackage{graphicx}

\begin{document}

   \title{On the circularly polarized optical emission from AE~Aquarii}

\author{N.R.\,Ikhsanov\inst1\fnmsep\inst2,
S.\,Jordan\inst3\fnmsep\inst4, and
N.G.\,Beskrovnaya\inst2\fnmsep\inst5}

  \offprints{N.R.~Ikhsanov \\ \email{ikhsanov@mpifr-bonn.mpg.de}}

   \institute{Max-Planck-Institut f\"ur Radioastronomie, Auf dem
              H\"ugel 69, D-53121 Bonn, Germany
              \and
              Central Astronomical Observatory of the Russian
              Academy of Science at Pulkovo, Pulkovo 65--1, 196140
              Saint-Petersburg, Russia
              \and
              Institut f\"ur Theoretische Physik und Astrophysik,
              Universit\"at Kiel, D-24098 Kiel, Germany
              \and
             Institut f\"ur Astronomie und Astrophysik, Sand 1,
             D-72076 T\"ubingen, Germany, \\
             \email{jordan@astro.uni-tuebingen.de}
              \and
             Isaac Newton Institute of Chile, St.Petersburg Branch.
}

   \date{Received 19 November 2001 / Accepted 17 January 2002}

\authorrunning{N.R.\,Ikhsanov \e}

\abstract{The reported nightly mean value of the circular
polarization of optical emission observed from the close binary
system \a is $0.06\% \pm 0.01\%$. We discuss the possibility that
the observed polarized radiation is emitted mainly by the white
dwarf or its vicinity. We demonstrate  that this hypothesis is
rather unlikely since the contribution of the white dwarf to the
optical radiation of the system is too small. This indicates that
the polarimetric data on \a cannot be used for the evaluation of
the surface magnetic field strength of the white dwarf in this
system. }

   \maketitle

   \section{Introduction}

Measurements of circularly polarized emission from the close
binary system \a have been reported by three independent groups of
authors. Cropper (\cite{c86}) estimated the circular polarization
in the optical to be at the level of $0.05\% \pm 0.01\%$. Similar
value (0.06\%) have been derived by Stockman \e (\cite{stock92}).
Beskrovnaya \e (\cite{bibs}) reported a value for the nightly mean
of the circular polarization in the $V$ and $V+R$ passbands to be
$p_{\rm mean}=0.06\% \pm 0.01\%$.

The possible origin of the circularly polarized emission from \a
was first discussed by Bastian \e (\cite{bdc88}) in the framework
of the {\it oblique rotator} model (Patterson \cite{p79}). Within
this model (which was widely accepted before 1994), \a is
considered as an {\it intermediate polar} in which the radiation
of the white dwarf is powered by the accretion of plasma onto its
surface with the rate of $\sim 10^{16}\,{\rm g\,s^{-1}}$. Assuming
the observed polarized emission to be generated in the shock at
the base of the accretion column due to the cyclotron mechanism,
Bastian \e (\cite{bdc88}) estimated the magnetic field strength of
the white dwarf to be in excess of $10^6$\,G.

This result, however, turned out to be one of the first serious
arguments against the accretion-powered white dwarf model of
AE~Aqr: If the surface field strength of the white dwarf in \a is
indeed $B_* \ga$\,1\,MG, its magnetospheric radius,
     \be\label{rm}
\r \simeq 7.7\,10^9\ \kappa_{0.5} \dot{M}_{16}^{-2/7}
M_{0.8}^{-1/7} R_{8.8}^{12/7} \left(\frac{B_*}{10^6\,{\rm
G}}\right)^{4/7} {\rm cm},
     \ee
is larger than the corotation radius,
   \be\label{rcor}
R_{\rm cor} = 1.5\,10^9 M_{0.8}^{1/3} P_{33}^{2/3}\,{\rm cm},
   \ee
by more than a factor of five. Here $\dot{M}_{16}$ is the mass
accretion rate expressed in units of $10^{16}\,{\rm g\,s^{-1}}$,
$\kappa_{0.5}=\kappa/0.5$ is  accounting for the geometry of
the accretion flow\footnote{The value of this
parameter lies within the interval $0.5 \la \kappa \la 1$, where
$\kappa = 0.5$ corresponds to the disk geometry of the accretion flow
and $\kappa = 1 $ -- to the spherical geometry (see e.g. Ghosh \& Lamb
\cite{gl79})}, and $M_{0.8}$, $R_{8.8}$ and $P_{33}$ are the mass,
radius, and the spin period of the white dwarf expressed in units of
$0.8 M_{\sun}$, $10^{8.8}$\,cm and 33\,s, respectively (for the system
parameters see Table~1 in Ikhsanov \cite{i00}). Under these
conditions, the white dwarf is in the centrifugal inhibition regime
and a steady accretion process onto its surface does not occur.

The hypothesis that the accretion power is not responsible for the
emission of the white dwarf in \a got serious grounds after 1994.
It was recognized that the intensity of radiation emitted from the
polar caps of the white dwarf does not correlate with the flaring
in the system (Eracleous \e \cite{erac94}). Furthermore, the X-ray
spectrum of \a is soft and essentially differs from the hard X-ray
spectra of intermediate polars (Clayton \& Osborne \cite{co95}).
Finally, the discovery of the rapid spin down of the white dwarf
(de~Jager \e \cite{jmor94}) and the conclusion that no developed
Keplerian accretion disk exists in the system (Wynn \e
\cite{wkh97}; Welsh \e \cite{whg98}) have left no doubts that \a
is not compatible with the {\it oblique rotator} model and that no
intensive plasma accretion onto the surface of the white dwarf
occurs (for discussion see Ikhsanov \cite{i01}). Therefore, the
basic assumptions used by Bastian \e (\cite{bdc88}) for the
interpretation of the circularly polarized emission of \a are
unacceptable and the origin of the polarized radiation in this
system remains an unresolved problem.

In this paper we are investigating the validity of the suggestion
that the observed circularly polarized radiation originates from
the white dwarf or its vicinity.

    \section{Contribution of the white dwarf to the circularly
polarized radiation}

It is presently well established that the optical radiation from
\a comes from at least three different sources: the
 K3--K5 main sequence companion, the white dwarf, and the
extended region  which is associated with the stream of material
lost by the normal companion and ejected  from the system due to
propeller action by the white dwarf. The latter source manifests
itself in the Balmer continuum and broad single-peaked emission
lines. This source is observed to be highly variable and is
assumed to be responsible for the unusual flaring behaviour of the
system in the optical/UV (Eracleous \& Horne \cite{eh96}, Wynn \e
\cite{wkh97}, Ikhsanov \cite{i00}). In contrast, investigations of
the normal companion (Welsh \e \cite{whg95}) and the white dwarf
(Eracleous \e \cite{erac94}) give no evidence for either
significant variations of their brightness or any correlation
between the intensity of radiation coming from these components
and the flaring in the system.

           \subsection{The contribution of the white dwarf to the
optical radiation of \a}

The visual light of \a is dominated mainly by the red dwarf: up to
95\% during quiescent state of the system (van Paradijs \e
\cite{pka89}, Bruch \cite{bruch91} and Welsh \e \cite{whg95}). The
contribution of the extended source to the optical radiation of \a
has been estimated by Eracleous \& Horne (\cite{eh96}) to amount
to about 3\%--4\% during quiescence and up to 40\% during the
flaring state of the system. Thus, the optical radiation from the
white dwarf proves to be a rather small fraction of the total
optical radiation detected from the system, namely, about 1\%--2\%
during the quiescent state and smaller than 1\% during the flaring
state.

This conclusion is in a good agreement with the model of the white
dwarf atmosphere reconstructed by Eracleous \e (\cite{erac94}) from
observations of \a with the Hubble telescope. Using the maximum
entropy method, the authors constrained the average surface temperature
of the white dwarf to the interval
 $10000\,{\rm K} \la T_{\rm int} \la 16000\,{\rm
K}$ and the temperature of the hot polar caps at the magnetic pole
regions to $24000\,{\rm K} \la T_{\rm max} \la 28000\,{\rm K}$
(the projected area of the hot spots is $4.3\,10^{16}\,{\rm
cm^2}$; the distance to \a is adopted to be 100\,pc). In the
framework of this model, the contribution of the hot polar caps to
the optical radiation of the system is about 0.1\%--0.2\% (that is
just the observed amplitude of the 33\,s coherent oscillations,
Patterson \cite{p79}, de~Jager \e \cite{jmor94}) and the total
contribution of the white dwarf is smaller than 2\%. This
indicates that the optical radiation coming from the hot polar
caps of the white dwarf is diluted by a factor of $k_{\rm cap} =
500 - 1000$ and the radiation of the white dwarf as a whole -- by
a factor of $k_{\rm wd} = 50 - 100$.

    \subsection{Pulse-averaged polarization}

It is important to note that the reported $p_{\rm mean}$ is the
average value of the circular polarization over a night. The
errors of the individual measurements are too large to be of any
significance (especially on the time scale of the white dwarf spin
period, $P=33$\,s). Moreover, the analysis of the polarimetric
data obtained by Beskrovnaya \e (\cite{bibs}) has shown no signs
of any periodicity. In this situation the idea that the polarized
radiation detected from \a comes mainly from the white dwarf or
its vicinity should be considered as an assumption rather than a
well justified assertion. If, nevertheless, we adopt this
assumption, it is necessary to take into account that the reported
degree of 0.06\% is not only the nightly mean but also the
pulse-averaged value. In this context, the polarization of
radiation coming from an individual magnetic pole should be larger
than the reported value $p_{\rm mean}$.

As the white dwarf rotates, its northern and southern magnetic
poles are successively seen. The contribution of each of the poles
to the system radiation depends on the configuration of the white
dwarf magnetic field, physical parameters of plasma in the pole
regions, and the system geometry. Most of these parameters have
been evaluated by Eracleous \e (\cite{erac94}) from the analysis
of the UV-spectrum and profiles of the 33\,s coherent
oscillations. According to their results, the longitudinal
difference between the hot polar caps is $180^{\degr}$, strongly
suggesting a dipolar geometry of the magnetic field of the white
dwarf. No evidence for any significant differences in the physical
parameters of the plasma situated at the northern and southern
poles has been found. Finally, assuming that the rotational axis
of the white dwarf is almost perpendicular to the orbital plane,
they evaluated the angle between the magnetic and the rotation
axes as $\beta\sim 75^{\degr}$ and the orbital inclination of the
system to be $i\simeq 60^{\degr}$ (see also Patterson \cite{p79}).

Under these conditions, radiation coming from the northern and southern
magnetic poles is circularly polarized to approximately the same
degree but in opposite directions:
$p_{\rm int}^{\rm n} \simeq -\,p_{\rm int}^{\rm s}$. This indicates
that the contributions of the northern and southern poles to the mean
value of the circular polarization observed from the system have
opposite signs and hence partly compensate each other. The value of
the non-compensated (i.e. pulse-averaged) polarization depends on
the system geometry.

\begin{figure}
\resizebox{\hsize}{!}{\includegraphics{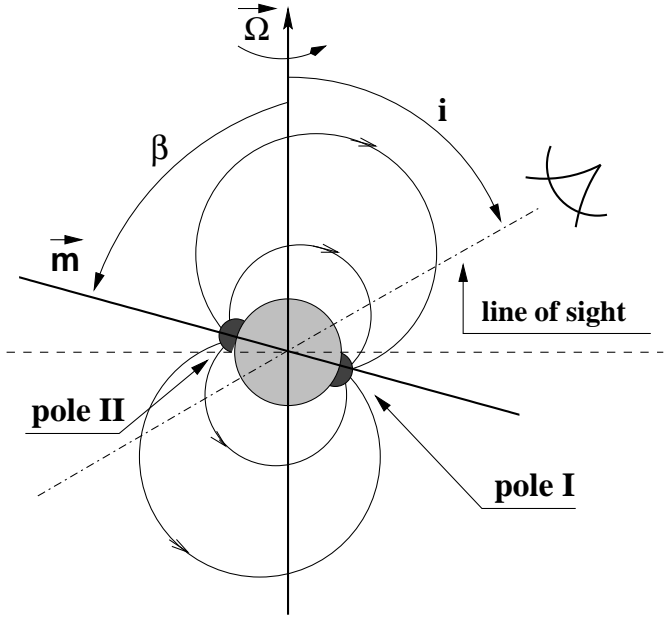}} \caption{The
relative geometry of the rotational and magnetic axes of the white
dwarf and the line of sight (see text)} \label{MS2107f1}
\end{figure}

The observed value of the polarization of an individual pole reaches
its maximum when the angle between the magnetic axis and the line of
sight is minimal and decreases to zero when this angle increases to
$90^{\degr}$ (complete cancellation). In our case the minimum angle
 between the line of sight
and the magnetic axis is $45^{\degr}$ for the pole\,I and is
$15^{\degr}$ for the pole\,II (see Fig.\,\ref{MS2107f1}). This
indicates that the contribution of the pole\,II to the mean
polarization detected from the system is larger than the
contribution of the pole\,I. In the first approximation, the lower
limit of the contribution of the pole\,I can be estimated by that
part of the contribution of the pole\,II when the latter is
observed within the following intervals of angles  $[-90^{\degr},
- 45^{\degr}]$ and $[90^{\degr}, 45^{\degr}]$. Hence, the
contribution of the pole\,II is not compensated only when the
angle between the magnetic axis and the line of sight is $\mid
\gamma \mid \la 45^{\degr}$. In other words, the uncompensated
circularly polarized emission is observable only during 1/4 of the
white dwarf spin period and hence the intrinsic polarization of
radiation coming from an individual pole is at least by a factor
of $k_{\rm pa}=4$ larger than the observed value.

   \subsection{Intrinsic polarization}

From the arguments of two previous subsections one can discard the
possibility that the hot polar caps,  situated in the magnetic
pole regions of the white dwarf, are the sources of the circularly
polarized optical emission detected from AE~Aqr: If the observed
polarized radiation were generated at the polar caps, the
intrinsic polarization of the source, $p_{\rm caps}=k_{\rm caps}
\times k_{\rm pa} \times p_{\rm mean}$, should be in excess of
100\% independent of the polarization mechanism\,!

There is, however, an alternative possibility. Namely, one can
assume that the circularly polarized radiation is generated in the
white dwarf atmosphere due to the linear and quadratic Zeeman
effect. In this case the effective area of the source is
significantly larger than the projected area of the hot polar
caps. Therefore the intrinsic polarization can be limited to the
interval $p_{\rm wd}=k_{\rm wd} \times k_{\rm pa} \times p_{\rm
mean} \ga 12\%$.

\begin{figure}
\resizebox{8cm}{10.5cm}{\includegraphics{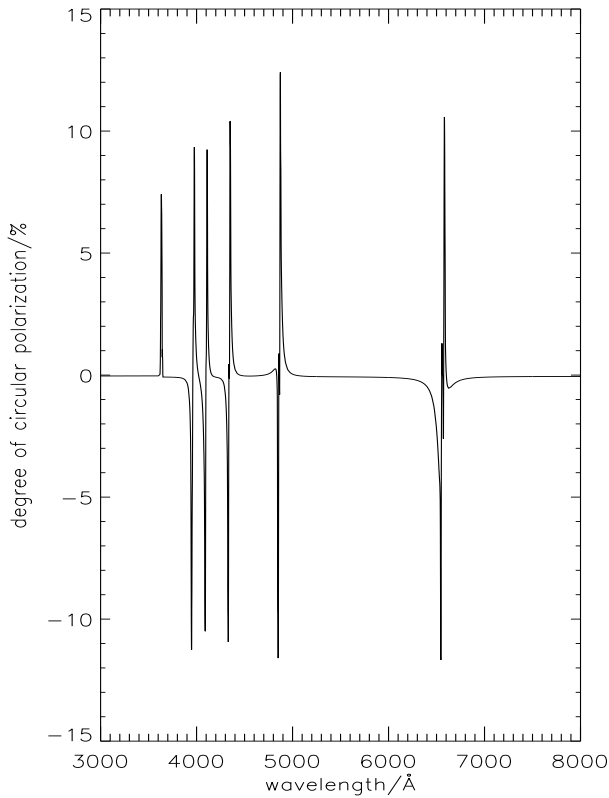}}
\caption{The calculated circular maximum possible polarization of
the white dwarf in AE~Aqr due to Zeeman effect. The model
parameters: $T_{\rm int} = 15000$\,K, $T_{\rm max} = 25000$\,K,
$\gamma=15^{\degr}$ and $B_*= 1$\,MG. Note that the polarization
in \a is diluted at least by a factor of 200 compared to this
prediction}
\label{MS2107f2}
\end{figure}

In order to test this possibility we estimated the expected value
of the circularly polarized emission from a non-accreting white
dwarf with the average surface temperature of 15\,000\,K, and two
hot spots ($T\sim25\,000$\,K) with the effective projected area of
$4.3\,10^{16}\,{\rm cm^2}$, situated at the magnetic pole regions.
The code developed by Jordan (\cite{jordan92}; see also Putney \&
Jordan \cite{pj95}) has been used to compute the polarization
degree at different wavelengths for an angle of  $15^{\degr}$
between the magnetic axis and the line of sight, corresponding to
the maximum value of the circular polarization. The results for
two values of the surface magnetic field strength, 1\,MG and
50\,MG, are plotted in Fig.\,\ref{MS2107f2} and \ref{MS2107f2},
respectively. As can be seen from these figures, the expected
maximum value of the polarization in the $V$ and  $V+R$ passbands
does not exceed 3\%, i.e. is at least by a factor of 4 smaller
than  required for the interpretation of the circular polarization
observed from the system (the average polarization calculated
within the interval \hbox{$\mid\,\gamma\,\mid <45^{\degr}$} is
smaller than the presented maximum value at least by a factor of
3).

\begin{figure}
\resizebox{8cm}{10.5cm}{\includegraphics{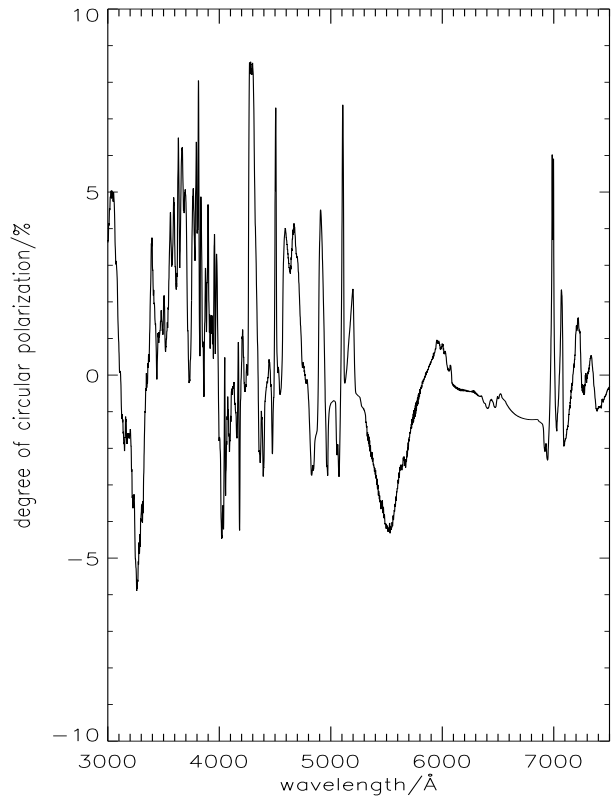}}
\caption{The same as Fig.\,\ref{MS2107f2} for $B_*= 50$\,MG.}
\label{MS2107f3}
\end{figure}

   \section{Discussion}

We have shown that the white dwarf cannot be
considered as the main source of the circularly polarized optical
emission detected from AE~Aqr. The basic argument is a very small
value of the white dwarf contribution to the visual light of
the system. As a consequence, the circularly polarized radiation
emitted by the white dwarf is significantly diluted and the
expected value of the pulse-averaged circular polarization proves
to be essentially smaller than the observed value.

Why was this argument  not taken into account in previous
investigations\,?
The reason was the widely accepted but wrong assumption that the system
 radiation is
powered by the accretion of material onto the white dwarf surface.
Within this assumption, the contribution of the white dwarf to the
system radiation proves to be overestimated by almost an order of
magnitude and thus, the problem discussed in this paper simply does
not arise. The incorrectness of this approach has been recognized
only a few years ago, mainly due to the discovery of the rapid
spin-down of the white dwarf (de~Jager \e \cite{jmor94}), detailed
investigations of the optical/UV properties of the 33\,s coherent
oscillations (Eracleous \e \cite{erac94}) and the reconstruction of
the diskless mass transfer picture (Wynn \e \cite{wkh97}).

One of the important consequences of our conclusion is that the
observed circularly polarized emission from \a cannot be used
to estimate the surface magnetic field strength of the white dwarf.
In the light of modern views on the system, neither the lower limit
suggested by Bastian \e (\cite{bdc88}) nor the upper limit given by
Stockman \e (\cite{stock92}) to the magnetic field strength of
the white dwarf can be used.
In this situation, the value of the white dwarf magnetic field should
be estimated using different methods (see e.g. Ikhsanov \cite{i98}).

Finally, we would like to note that except for the uncertain nightly mean
value, no information about the properties of the circularly
polarized emission from \a is currently available. In this situation,
the justification of any alternative ideas about the origin of the
polarized emission is very complicated and perhaps even impossible,
until highly time resolved  polarimetric observations with large
signal-to-noise ratio are performed.

\begin{acknowledgements}
We would like to thank Boris Gaensicke for useful comments. NRI
acknowledge the support of the Alexander von Humboldt Foundation
within the Long-term Cooperation Program. Work on magnetic white
dwarfs in Kiel was supported by the DFG under KO-738/7-1.
\end{acknowledgements}


\begin{thebibliography}{}
\bibitem[1988]{bdc88}
  Bastian T.S., Dulk G.A., Chanmugam G., 1988, ApJ 324, 431
\bibitem[1996]{bibs}
  Beskrovnaya N.G., Ikhsanov N.R., Bruch A., Shakhovskoy N.M., 1996,
        A\&A 307, 840
\bibitem[1991]{bruch91}
  Bruch A., 1991, A\&A 251, 59
\bibitem[1995]{co95}
  Clayton K.L., Osborne J.P., 1995, in ``Magnetic Cataclysmic Variables'',
     Eds. D.~Buckley and B.~Warner, ASP Conference Series 85, 379
\bibitem[1986]{c86}
  Cropper M., 1986, MNRAS 222, 225
\bibitem[1994]{jmor94}
  de~Jager O.C., Meintjes P.J., O'Donoghue D., Robinson E.L., 1994,
     MNRAS 267, 577
\bibitem[1996]{eh96}
  Eracleous M., Horne K., 1996, ApJ 471, 427
\bibitem[1994]{erac94}
  Eracleous M., Horne K., Robinson E.L., et~al., 1994, ApJ 433, 313
\bibitem[1979]{gl79}
  Ghosh P., Lamb F.K., 1979, ApJ 232, 259
\bibitem[1998]{i98}
  Ikhsanov N.R., 1998, A\&A 338, 521
\bibitem[2000]{i00}
  Ikhsanov N.R., 2000, A\&A 358, 201
\bibitem[2001]{i01}
  Ikhsanov N.R., 2001, A\&A 374, 1030
\bibitem[1992]{jordan92}
  Jordan S., 1992, A\&A 265, 570
\bibitem[1979]{p79}
  Patterson J., 1979, ApJ 234, 978
\bibitem[1995]{pj95}
  Putney A., Jordan S., 1995, ApJ 449, 863
\bibitem[1992]{stock92}
  Stockman H.S., Schmidt G.D., Berriman G., \e, 1992, ApJ 401, 628
\bibitem[1989]{pka89}
  van Paradijs J., Kraakman H., van Amerongen S., 1989, A\&AS 79,
     205
\bibitem[1995]{whg95}
  Welsh W.F., Horne K., \& Gomer R., 1995, MNRAS 275, 649
\bibitem[1998]{whg98}
  Welsh, W.F., Horne, K., Gomer, R., 1998, MNRAS 298, 285
\bibitem[1997]{wkh97}
  Wynn G.A., King A.R., Horne K., 1997, MNRAS 286, 436
\end{thebibliography}
\end{document}